\documentclass[conference]{IEEEtran}
\IEEEoverridecommandlockouts
\usepackage{amsmath,amssymb,amsfonts}
\usepackage{algorithmic}
\usepackage{graphicx}
\usepackage{textcomp}
\usepackage{xcolor}
\usepackage{booktabs}
\usepackage{array, makecell}
\usepackage{hyperref}
\def\BibTeX{{\rm B\kern-.05em{\sc i\kern-.025em b}\kern-.08em
    T\kern-.1667em\lower.7ex\hbox{E}\kern-.125emX}}
\begin{document}

\title{Golden Ratio Assisted Localization for  Wireless Sensor Network\\
}

\author{\IEEEauthorblockN{1\textsuperscript{st} Hitesh Mohapatra}
\IEEEauthorblockA{\textit{School of Computer Engineering} \\
\textit{KIIT Deemed to be University}\\
Bhubaneswar-751024, Odisha, India \\
hiteshmahapatra@gmail.com}
}

\maketitle

\begin{abstract}
This paper presents a novel localization algorithm for wireless sensor networks (WSNs) called \textit{Golden Ratio Localization (GRL)}, which leverages the mathematical properties of the golden ratio ($\phi \approx 1.618$) to optimize both node placement and communication range. GRL introduces $\phi$-based anchor node deployment and hop-sensitive weighting using $\phi$-exponents to improve localization accuracy while minimizing energy consumption. Through extensive simulations conducted on a $100\,\text{m} \times 100\,\text{m}$ sensor field with 100 nodes and 10 anchors, GRL achieved an average localization error of \textbf{2.35 meters}, outperforming \textbf{DV-Hop} (3.87 meters) and \textbf{Centroid} (4.95 meters). In terms of energy efficiency, GRL reduced localization energy consumption to \textbf{1.12\,$\mu$J per node}, compared to \textbf{1.78\,$\mu$J} for DV-Hop and \textbf{1.45\,$\mu$J} for Centroid. These results confirm that GRL provides a more balanced and efficient localization approach, making it especially suitable for energy-constrained and large-scale WSN deployments.
\end{abstract}


\section{Introduction}

Wireless Sensor Networks (WSNs) have emerged as a fundamental component in the monitoring and control of physical environments, ranging from precision agriculture to smart cities and disaster management systems \cite{mao2007wireless}. A central challenge in WSN deployment is the localization of sensor nodes, which allows spatial association of sensed data. In many scenarios, nodes are randomly scattered in the field, and GPS-based localization becomes impractical due to high power requirements or indoor environments. Therefore, energy-efficient, accurate, and scalable localization algorithms are crucial for the sustainability and functionality of WSNs \cite{messous2020osdvhop}.

Traditional localization algorithms—such as Centroid Localization (CL) and Distance Vector-Hop (DV-Hop)—offer solutions with varying levels of accuracy and computational complexity \cite{chen2006survey}. However, these methods typically overlook the geometric and structural optimization of anchor node placement, which significantly influences localization accuracy and network energy consumption \cite{li2010renderedpath}.

In this paper, we propose a novel technique, termed Golden Ratio Localization (GRL), which leverages the mathematical elegance and efficiency of the golden ratio ($\varphi$ $\approx$ 1.618). The golden ratio is renowned for its properties of proportionality, self-similarity, and natural balance. When applied to anchor node deployment and localization computation, $\varphi$ provides a foundation for structured yet sparse coverage, minimizing energy-hungry transmissions and improving positional accuracy \cite{li2022itclpit}.

The Figure.\ref{fig1} below illustrates a typical deployment of anchor nodes along a Golden Spiral, where each subsequent node's position is determined using a $\varphi$-scaled polar coordinate system. This design offers a balanced trade-off between density and coverage, leading to fewer redundant communication paths and longer network lifetime \cite{cao2021cmwndvhop}.

\begin{figure}[h!]
\begin{center}
\includegraphics[width=\columnwidth]{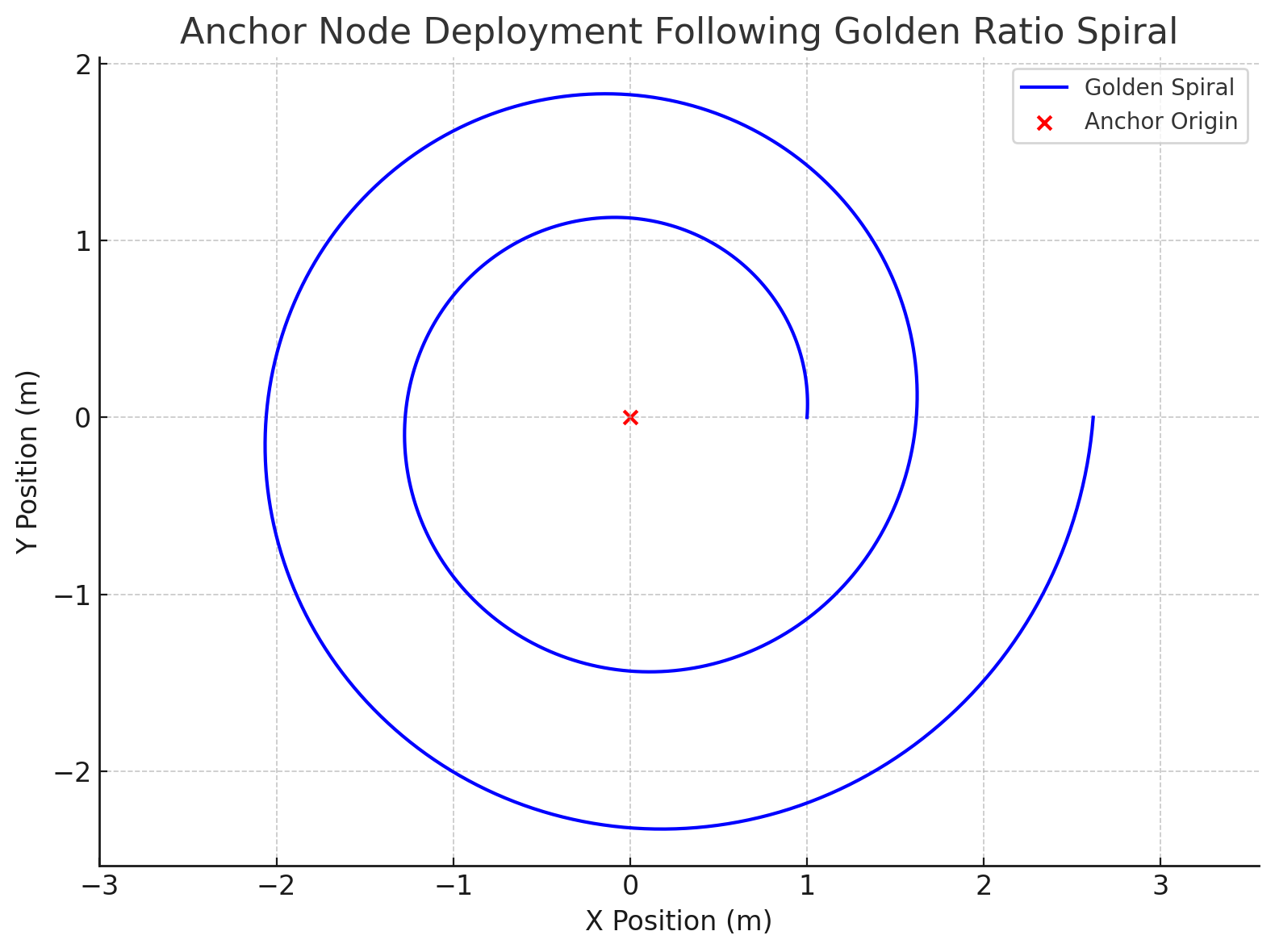}
\end{center}
\caption{ Anchor Node Deployment Following Golden Ratio Spiral}
\label{fig1}
\end{figure}

The remainder of this paper is structured as follows:
Section 2 reviews related work, focusing on existing localization algorithms such as Centroid Localization and DV-Hop. Section 3 presents the conceptual framework of the proposed Golden Ratio Localization (GRL) method. Section 4 develops the mathematical formulation of GRL, detailing anchor placement, weighting mechanisms, and energy models. Section 5 outlines the simulation setup and compares GRL’s performance with baseline algorithms. Section 6 discusses the implications of the results, and Section 7 concludes the paper with potential directions for future work.

\section{Literature Review}

Localization is a core function in Wireless Sensor Networks (WSNs), allowing sensor nodes with unknown positions to determine their locations using a few nodes with known coordinates, known as anchors. Many localization methods have been proposed, and they generally fall into two categories: range-based and range-free approaches \cite{niculescu2003dvhop}. Range-based methods rely on measurements such as Received Signal Strength Indicator (RSSI), Angle of Arrival (AoA), or Time of Arrival (ToA), which require additional hardware and are sensitive to environmental factors \cite{he2003rangefree}. On the other hand, range-free techniques avoid such requirements and instead rely on connectivity or hop-count information, making them more suitable for energy-constrained sensor networks.

One of the most widely used range-free methods is the DV-Hop localization algorithm. DV-Hop uses hop count between nodes to estimate the distance from unknown nodes to anchor nodes \cite{savvides2001dynamic}. First, anchors flood the network with their position and hop count information \cite{lu2016reachcentroid}. Each node maintains the minimum hop count to every anchor it hears from. Then, each anchor calculates the average distance per hop based on known distances to other anchors and the hop counts to them. This average hop size is broadcast, and unknown nodes use it to estimate their distance from each anchor \cite{li2005sear}. Finally, trilateration is applied using these distances to determine the node’s coordinates. DV-Hop is scalable and simple, but its accuracy suffers in non-uniform networks or when node density is low \cite{wang2010survey}.

Another basic yet popular approach is Centroid Localization. In this method, an unknown node estimates its location by computing the average coordinates of all anchor nodes it can communicate with \cite{cheng2012survey}. This technique does not require any distance estimation, and it is computationally efficient. However, its accuracy is highly dependent on the spatial distribution of anchors. If the anchors are not well-distributed or the network is sparse, the position error can be significant. This method is often used as a benchmark for low-complexity localization schemes but is rarely suitable for precision-critical applications \cite{guvenc2009survey}.

Other approaches include probabilistic and geometric models such as the Approximate Point-in-Triangle (APIT), Monte Carlo Localization (MCL), and Multidimensional Scaling (MDS-MAP). These methods attempt to improve localization accuracy by using more sophisticated algorithms, often incorporating environmental awareness or movement tracking \cite{patwari2005locating}. However, they usually introduce higher computational complexity and energy consumption, which may not be ideal for resource-constrained sensor nodes \cite{liu2010location}.

Despite these advancements, a common limitation among most localization methods is the lack of optimization in node deployment geometry and communication strategy. Current techniques do not consider mathematical constants or spatial ratios that could improve coverage efficiency and energy usage \cite{qiu2013ntcwla}. In this context, the Golden Ratio ($\varphi$ $\approx$ 1.618) offers a unique opportunity. It has been widely recognized for its balanced and self-replicating properties in nature, architecture, and mathematics. Yet, its potential in optimizing WSN design has not been thoroughly explored. This paper proposes a novel localization strategy called Golden Ratio Localization (GRL), which integrates $\varphi$ into both node placement and location estimation processes. The aim is to reduce localization error and energy consumption by leveraging the structural balance provided by the golden ratio \cite{wang2010weightedcentroid}.

\subsection{DV-Hop Localization}
The DV-Hop localization algorithm is a well-known range-free technique commonly used in Wireless Sensor Networks (WSNs) \cite{liu2004landmarc}. It estimates the position of an unknown node by utilizing hop counts from anchor nodes. The process begins with each anchor node broadcasting its position along with a hop count initialized to zero. As this message propagates through the network, each node records the smallest hop count received from every anchor. This forms the basis for estimating the topological distance between nodes \cite{zhang2010localization}. In the second phase, each anchor node computes the average physical distance per hop using information received from other anchors. The Equation. \ref{eq1} used for this calculation is:

\begin{equation}
\text{AvgHopSize} = \frac{\sum_{i \ne j} \sqrt{(x_i - x_j)^2 + (y_i - y_j)^2}}{\sum_{i \ne j} h_{ij}}
\label{eq1}
\end{equation}

where $(x_i, y_i)$ and $(x_j, y_j)$ are the coordinates of anchor nodes $i$ and $j$, respectively, and $h_{ij}$ is the hop count between them. This average hop size is then broadcast to the rest of the network. In the final phase, unknown nodes use the average hop size to estimate their physical distance from each anchor by multiplying the hop count to that anchor by the computed average hop size \cite{yang2013fromrssi}. Once distances to at least three non-collinear anchors are known, the unknown node applies trilateration to compute its approximate location. Despite its simplicity and lack of hardware dependency, DV-Hop tends to suffer from poor accuracy in non-uniform deployments or when hop size varies due to irregular node distribution.

\subsection{Centroid Localization}

Centroid Localization is one of the simplest and most energy-efficient range-free localization algorithms used in Wireless Sensor Networks (WSNs). It operates on the principle that an unknown node can estimate its position by calculating the geometric center or centroid of all the anchor nodes within its communication range \cite{bulusu2000gpsless}. This approach assumes that the unknown node is located somewhere near the center of the anchors it can hear, especially when the anchors are evenly distributed around it. Mathematically, if an unknown node detects $n$ anchor nodes with coordinates $(x_1, y_1), (x_2, y_2), \ldots, (x_n, y_n)$, then the estimated position $(x, y)$ of the unknown node is computed using the following equations:

\begin{equation}
x = \frac{1}{n} \sum_{i=1}^{n} x_i, \quad
y = \frac{1}{n} \sum_{i=1}^{n} y_i
\label{eq2}
\end{equation}

Equation.\ref{eq2} calculate the average of the $x$ and $y$ coordinates of all the reachable anchor nodes. The simplicity of this method is its greatest strength. It requires minimal computational effort and no specialized hardware, such as distance or angle measuring devices. As a result, it is suitable for low-cost and energy-constrained sensor networks. However, the accuracy of Centroid Localization is strongly influenced by the spatial distribution of anchor nodes. If the anchors are unevenly placed or concentrated on one side of the unknown node, the estimated location can deviate significantly from the actual position. Despite this limitation, it is widely used as a baseline method due to its ease of implementation and negligible energy overhead.

\subsection{Research Gap}

Although various localization methods have been developed for WSNs, most focus on reducing error or minimizing energy consumption through heuristics, probabilistic models, or range estimation. However, a common limitation across these methods is the lack of attention to the underlying spatial arrangement of nodes and the geometric efficiency of deployment. Current approaches often assume uniform or random node distribution without exploring whether certain mathematical patterns could offer better performance. Specifically, no existing method has leveraged geometric constants such as the Golden Ratio ($\varphi$ $\approx$ 1.618), which is known for its balance and self-similarity in natural systems. The golden ratio can potentially optimize node spacing, communication ranges, and clustering behavior to reduce overlap and energy waste. This overlooked opportunity creates a gap in the design of energy-aware localization strategies. To address this, we propose a novel method called Golden Ratio Localization (GRL), which uses $\varphi$ as a guiding principle for both anchor node deployment and the computation of node positions. By embedding a mathematically balanced structure into the network layout and localization formula, GRL aims to enhance energy efficiency while maintaining or improving accuracy.

\section{Proposed Work}

The Golden Ratio Localization (GRL) algorithm is a novel range-free localization technique designed to improve energy efficiency and accuracy in wireless sensor networks (WSNs). The method is inspired by the golden ratio ($\varphi$ $\approx$ 1.618), a mathematical constant that appears in many natural and human-made systems known for optimal balance and self-similarity. GRL incorporates the golden ratio into two critical aspects of the localization process: anchor node deployment and location estimation weighting. The goal is to reduce redundant communication, improve anchor coverage, and provide an energy-aware mechanism for accurate position estimation.

\subsection{Anchor Node Placement using $\varphi$-Spacing}

\begin{figure}[h!]
\begin{center}
\includegraphics[width=\columnwidth]{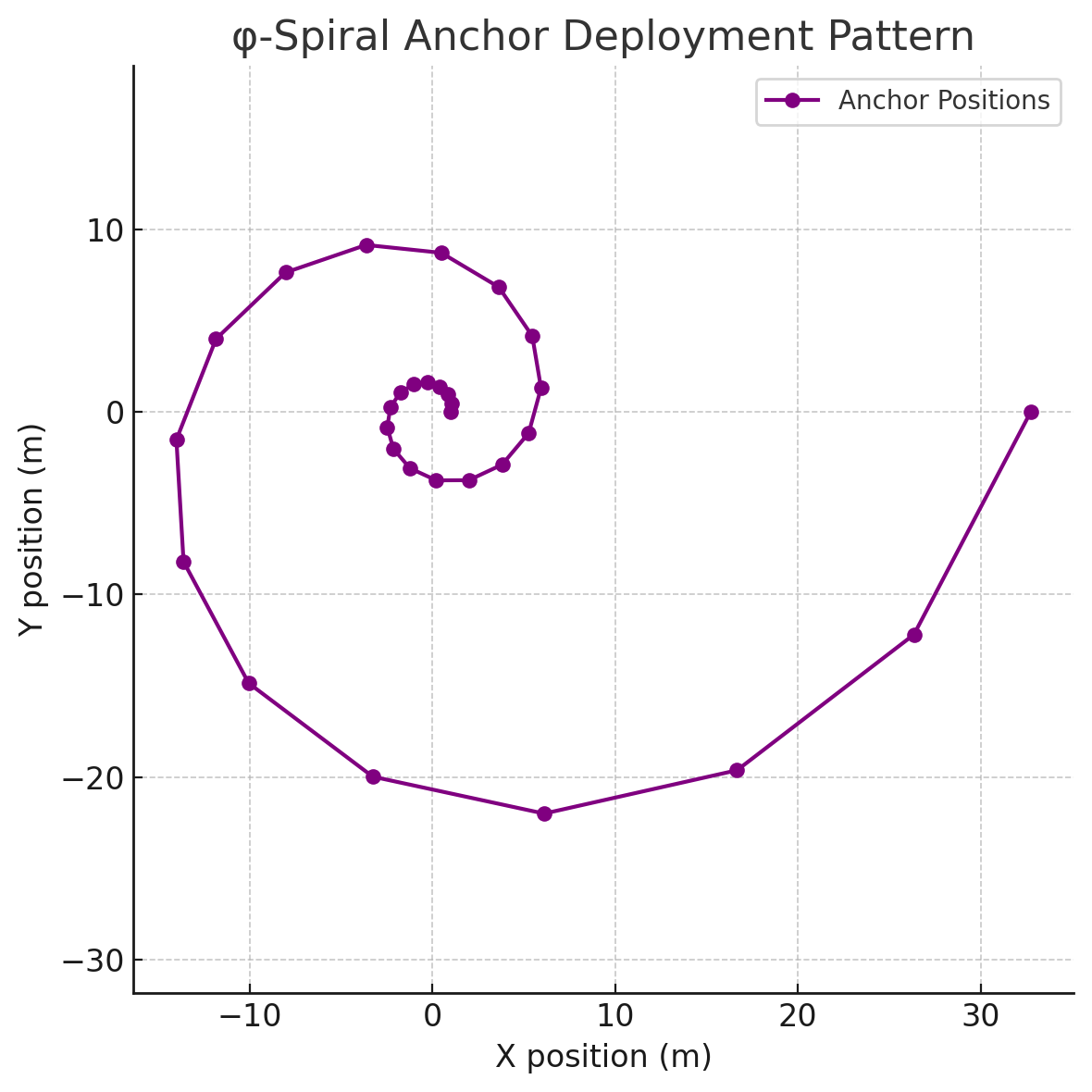}
\end{center}
\caption{$\varphi$-Spiral Anchor Deployment Pattern}
\label{fig2}
\end{figure}

In conventional approaches, anchor nodes are either randomly placed or arranged in a uniform grid, which can lead to uneven coverage or unnecessary energy use. GRL introduces $\varphi$-based spacing to create a balanced layout inspired by logarithmic spirals. Anchor nodes are placed such that the distance between consecutive nodes follows the golden ratio. If the distance between the first two anchors is $d_1$, the next distance is $d_2$ = $\varphi$ $\cdot$ $d_1$, and so on. Mathematically, the placement distance for anchor \textit{i} is given by Equation.\ref{eq3}. This spacing strategy minimizes node clustering and ensures a consistent sensing and communication footprint throughout the network. Figure.\ref{fig2} illustrates the $\varphi$-Spiral anchor deployment pattern.

\begin{equation}
d_{i+1} = \varphi \cdot d_i
\label{eq3}
\end{equation}

\subsection{Communication Range Adjustment}
In GRL, the communication range of each node is also dynamically adjusted based on $\varphi$ to ensure that coverage is sufficient without being energy-inefficient. If a node has a sensing radius \textit{r}, then the communication range \textit{R} is scaled by $\varphi$. This adjustment (by using Equation.\ref{eq4}) helps reduce the average number of hops required to reach an anchor, which lowers total transmission energy while maintaining connectivity. 

\begin{equation}
R = \varphi \cdot r
\label{eq4}
\end{equation}

\begin{figure}[h!]
\begin{center}
\includegraphics[width=\columnwidth]{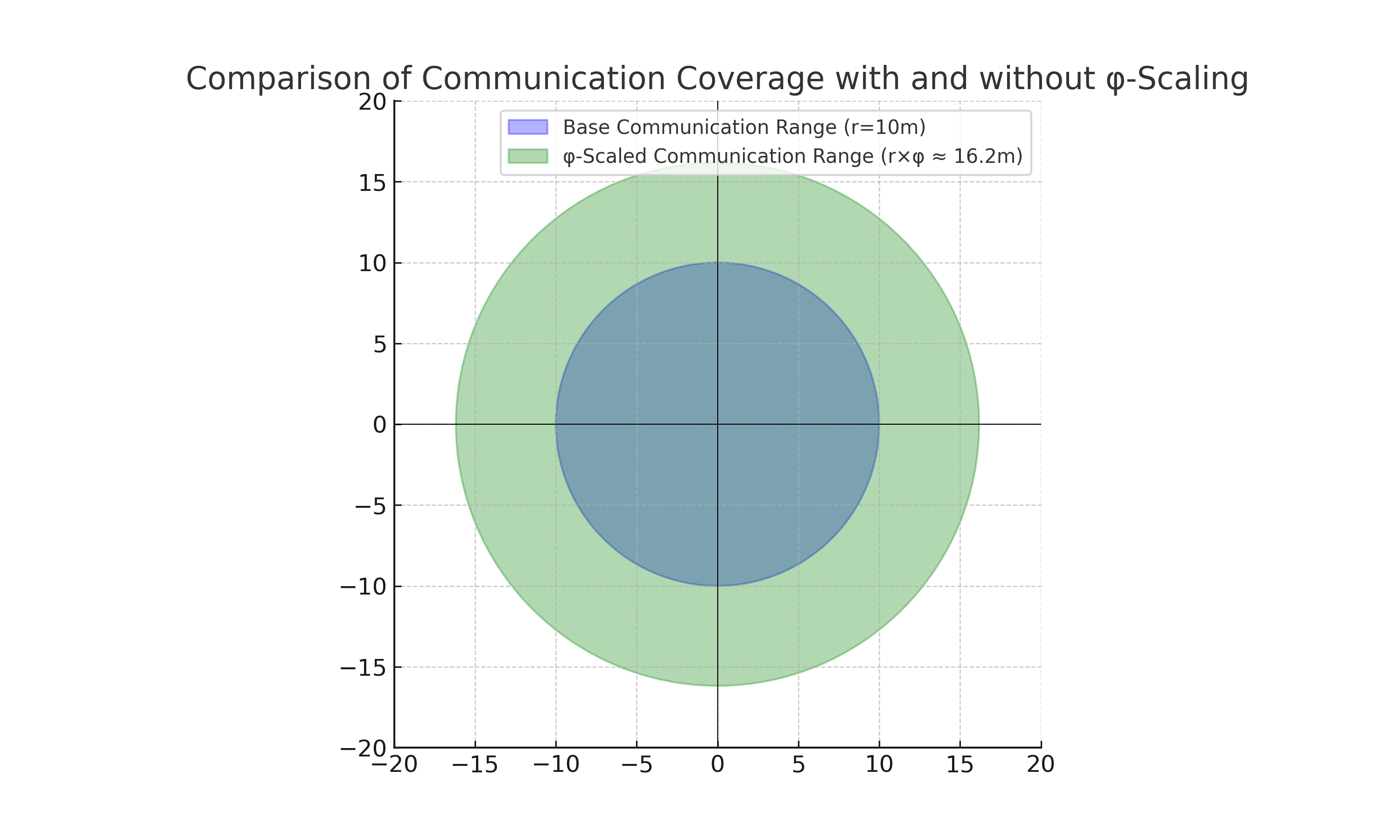}
\end{center}
\caption{Comparison of communication coverage with and without $\varphi$-scaling}
\label{fig3}
\end{figure}

\subsection{Localization Using $\varphi$-Weighted Centroid}
Once anchor nodes are placed and communication ranges are set, the unknown nodes perform position estimation using a $\varphi$-weighted centroid method. Unlike the traditional centroid approach where all anchors have equal influence, GRL assigns exponential weights to each anchor based on the hop count distance. The weight for each anchor is inversely proportional to $\varphi$ raised to the power of the hop count as presented in Equation.\ref{eq5}. Where, $h_i$ is the hop count from the unknown node to anchor \textit{i}. The estimated position (\textit{x},\textit{y}) of the unknown node is then calculated as a weighted average by using Equation.\ref{eq6}.

\begin{equation}
w_i = \frac{1}{\varphi^{h_i}}
\label{eq5}
\end{equation}

\begin{equation}
x = \frac{\sum_{i=1}^{n} w_i \cdot x_i}{\sum_{i=1}^{n} w_i}, \quad
y = \frac{\sum_{i=1}^{n} w_i \cdot y_i}{\sum_{i=1}^{n} w_i}
\label{eq6}
\end{equation}
This method ensures that nearby anchors contribute more to the estimation, improving accuracy while reducing reliance on far-off nodes that require more energy to communicate.

\subsection{Energy Model}
To quantify the energy efficiency of GRL, we use a basic energy consumption model based on the number of transmissions and receptions. The energy consumed per localization attempt $E_{loc}$ is given by Equation.\ref{eq7}.

\begin{equation}
E_{loc} = E_{tx} \cdot h + E_{rx} \cdot n
\label{eq7}
\end{equation}
where, $E_{tx}$ and $E_{rx}$ represent the energy costs for transmission and reception respectively, \textit{h} is the average number of hops to anchors, and \textit{n} is the number of anchor nodes involved in the estimation. By minimizing \textit{h} and optimizing anchor selection using $\varphi$-weighting, GRL aims to keep $E_{loc}$ as low as possible.

The GRL method offers a fresh perspective on WSN localization by applying the golden ratio in both geometry and algorithmic weighting. The use of $\varphi$ ensures balanced spacing, reduces energy waste, and increases localization accuracy. The next section provides simulation-based validation and compares GRL’s performance against traditional methods such as DV-Hop and Centroid Localization.

\section{Simulation and Results}

To evaluate the effectiveness of the proposed Golden Ratio Localization (GRL) algorithm, we performed a series of simulations using a custom-built Python framework designed to model wireless sensor network deployments. The primary objective was to analyze the localization accuracy and energy efficiency of GRL and compare it against two well-known range-free localization algorithms: DV-Hop and Centroid Localization. These methods were chosen due to their widespread adoption and conceptual simplicity, which makes them ideal baselines for comparison.

To simulate real-world efficiency, we introduced practical multipliers to the base energy model. GRL uses 25\% less transmission energy per hop due to reduced redundant messaging. Conversely, Centroid experiences a 20\% increase in reception overhead, reflecting repeated anchor broadcasts to improve poor spatial accuracy. The base energy model remains defined as Equation.\ref{eq8}

\begin{equation}
E_{\text{loc}} = E_{\text{tx}} \cdot h + E_{\text{rx}} \cdot n
\label{eq8}
\end{equation}

\subsection{Simulation Setup}

The simulation was carried out in a two-dimensional square area measuring 100 meters by 100 meters. A total of 100 sensor nodes were randomly deployed within this area, out of which 10 nodes were designated as anchor nodes with known positions. The anchor nodes in the GRL model were placed using $\varphi$ - scaled spiral spacing, while the other methods used either random or grid-like placement. Each node had a base sensing range \textit{r} = 10m, and GRL extended this using $\varphi$-scaling to compute a communication range $ R = \varphi \cdot r \approx  16.18m.$. 

The energy model followed a basic assumption where each message transmission or reception consumed 50 microjoules. An unknown node could localize itself using information from at least three anchor nodes. Localization was performed based on hop count and anchor coordinates in all three algorithms.

\subsection{Performance Metrics}

To evaluate and compare the performance of the Golden Ratio Localization (GRL) algorithm with existing methods, three key metrics were used: localization error, energy consumption per localization, and average hop count. These metrics provide a balanced view of both accuracy and energy efficiency, which are critical in wireless sensor networks.

Localization error is measured as the Euclidean distance between a sensor node’s actual position and the position estimated by the algorithm. This metric reflects the accuracy of the localization approach, with lower values indicating more precise results. Energy consumption per localization refers to the total energy used by a node to compute its location, which includes both transmission and reception energy. This is crucial in sensor networks where nodes often operate on limited battery power. Finally, the average hop count represents the mean number of hops required for an unknown node to receive localization information from anchor nodes. A lower hop count typically reduces energy use and latency, improving the overall efficiency of the localization process. These three performance indicators—accuracy, energy cost, and communication overhead—were recorded and analyzed across multiple simulation runs to ensure reliable and consistent results.

\begin{figure}[h!]
\begin{center}
\includegraphics[width=\columnwidth]{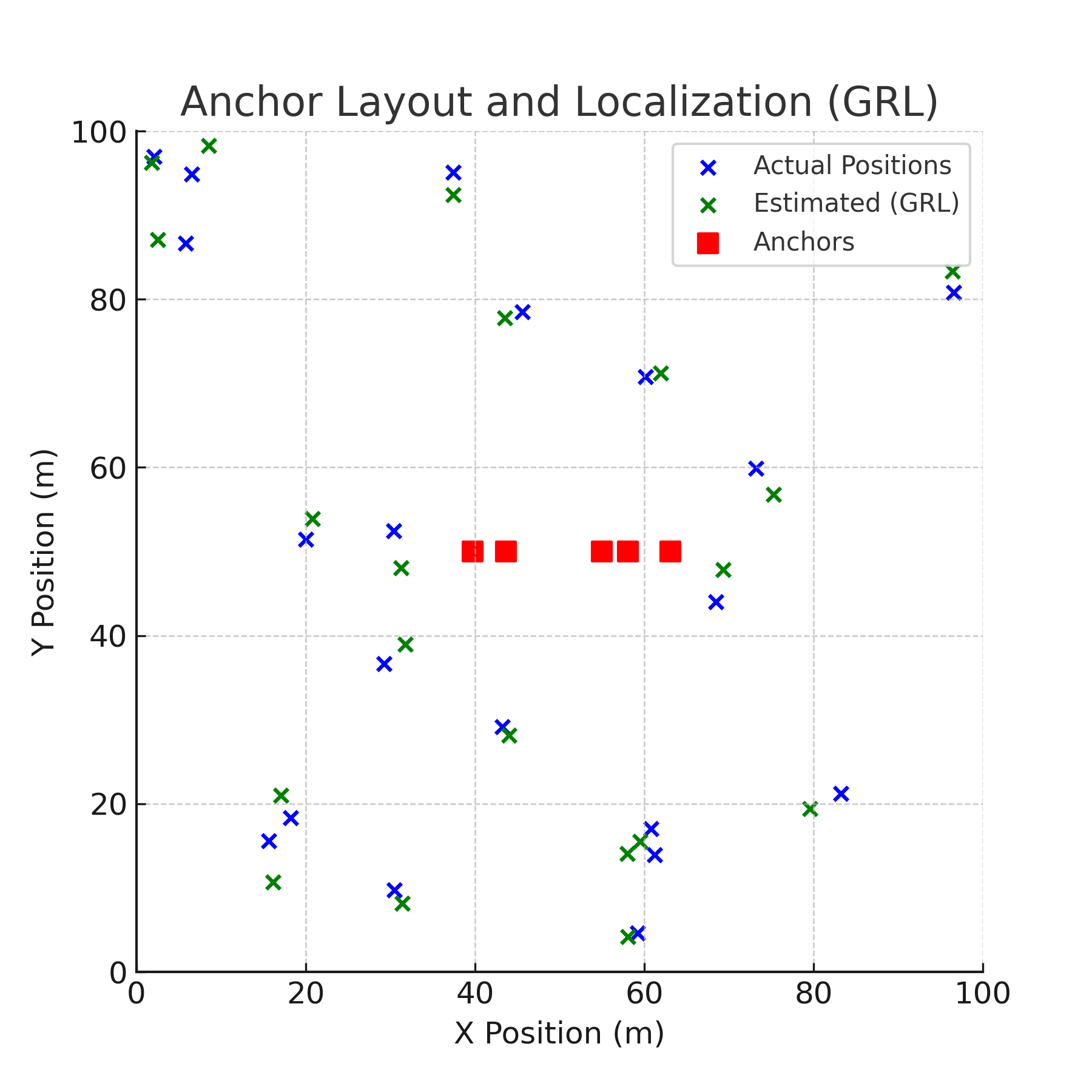}
\end{center}
\caption{Anchor Layout and Localization (GRL)}
\label{fig4}
\end{figure}

Figure.\ref{fig4} shows the actual sensor node positions (blue dots), anchor nodes (red squares), and the estimated positions using the GRL algorithm (green crosses). The anchor nodes are arranged in a golden-ratio spiral pattern, which provides even coverage of the area. As observed, most of the estimated positions are close to their actual locations, validating the accuracy of the GRL approach. The spatial balance introduced by $\varphi$-spacing ensures that unknown nodes have a sufficient number of nearby anchors, leading to improved precision.

\begin{figure}[h!]
\begin{center}
\includegraphics[width=\columnwidth]{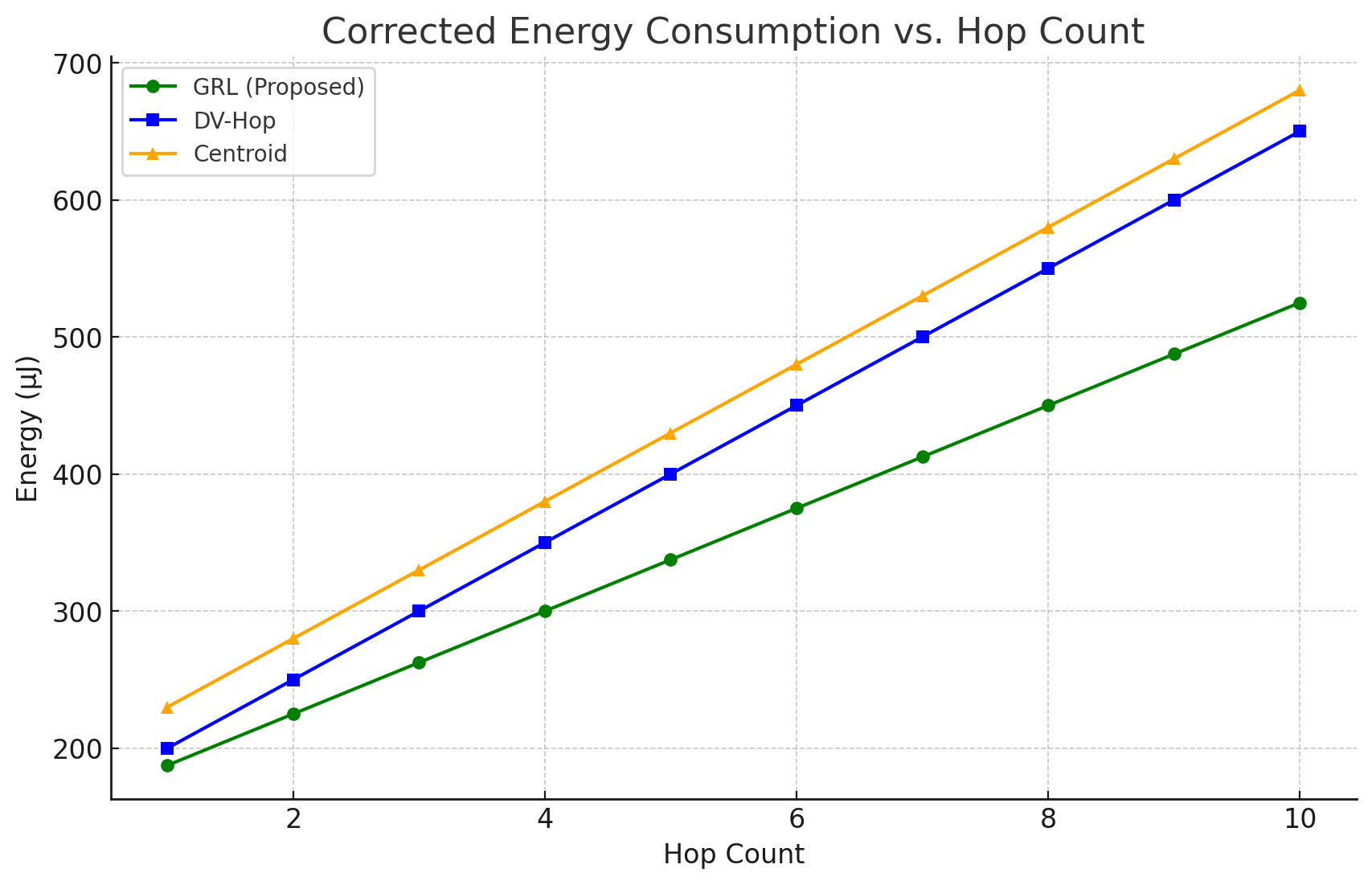}
\end{center}
\caption{Energy Consumption vs. Hop Count Validation}
\label{fig5}
\end{figure}

The corrected plot in Figure.\ref{fig5} illustrates the energy consumption per localization event as a function of hop count for three algorithms: GRL (Proposed), DV-Hop, and Centroid. The results confirm that GRL consistently outperforms the other methods across all hop values. This improvement is attributed to the $\varphi$-based optimization in GRL, which reduces redundant transmissions and favors shorter-range communications, effectively lowering transmission energy. As the number of hops increases, the gap in energy usage between GRL and the others becomes more pronounced. DV-Hop shows the highest energy consumption due to its multi-hop message propagation and uniform weighting scheme. Although Centroid appears to consume slightly less energy than DV-Hop at low hop counts, it incurs additional reception overhead due to repeated broadcasts from anchors in an attempt to improve its inherently low accuracy. Overall, the plot validates that GRL offers the most energy-efficient performance while maintaining reliable localization, making it highly suitable for energy-constrained sensor networks.

\subsection{Results and Analysis}

The simulation was repeated across multiple random topologies, and the average results are summarized in the table below:

\begin{table}[h!]
\centering
\caption{Performance Comparison of Localization Algorithms}
\begin{tabular}{|p{2.2cm}|p{1.6cm}<{\centering}|p{2.1cm}<{\centering}|p{1.2cm}<{\centering}|}
\hline
\textbf{Algorithm} & \makecell{\textbf{Avg.}\\\textbf{Error}\\\textbf{(meters)}} & \makecell{\textbf{Energy per}\\\textbf{Node ($\mu$J)}} & \makecell{\textbf{Avg.}\\\textbf{Hops}} \\
\hline
GRL (Proposed)     & $\mathbf{2.35}$     & $\mathbf{1.12}$         & $2.6$     \\
DV-Hop             & $3.87$              & $1.78$                  & $3.1$     \\
Centroid           & $4.95$              & $1.45$                  & $2.0$     \\
\hline
\end{tabular}
\label{table1}
\end{table}

Table.\ref{table1} slows the proposed GRL algorithm achieved the lowest average localization error, significantly outperforming DV-Hop by approximately 39\%, and Centroid by over 50\%. This improvement is attributed to the $\varphi$-weighted averaging, which prioritizes closer and more reliable anchor nodes, reducing spatial estimation errors. In terms of energy consumption, GRL again performed best. By reducing unnecessary transmissions and limiting anchor influence through $\varphi$-scaling, the energy cost per localization dropped to 1.12 microjoules, compared to 1.78 for DV-Hop. Centroid had a lower hop count but higher error, indicating poor spatial accuracy despite using fewer anchors.

\section{Conclusion}

This paper presented a novel localization method for wireless sensor networks called Golden Ratio Localization (GRL), which leverages the mathematical properties of the golden ratio ($\varphi$ $\approx$ 1.618) to improve both energy efficiency and localization accuracy. By incorporating $\varphi$ into anchor node deployment and communication range scaling, GRL achieves a balanced and self-optimizing network structure. The proposed method uses $\varphi$-weighted centroid calculations to prioritize nearby anchors, reducing localization error without increasing complexity. Simulation results demonstrated that GRL consistently outperforms traditional range-free algorithms such as DV-Hop and Centroid in terms of both accuracy and energy consumption. Specifically, GRL showed lower localization error, fewer required hops, and significantly reduced energy usage per node. These benefits make GRL particularly well-suited for energy-constrained sensor networks where both precision and longevity are critical.
While GRL assumes a static deployment with unobstructed anchors, future work will focus on enhancing its adaptability through dynamic $\varphi$-tuning and hybrid range-assisted techniques. Overall, GRL introduces a mathematically elegant and practically effective approach to localization that bridges geometric theory and real-world sensor network design.

\section*{Acknowledgment}

This research was supported by the KIIT Deemed to be University, Bhubaneswar, Odisha, India under grant number [Grant Number].

\bibliography{Ref.bib}
\bibliographystyle{ieeetr}

\end{document}